\documentclass[a4paper,11pt]{article}
\usepackage{jcappub} % for details on the use of the package, please see the JINST-author-manual
\usepackage{aas_macros} %bibliography macros
\newcommand{\added}[1]{#1}
\arxivnumber{2607.06737} % Only if you have one
\makeatletter
\gdef\@fpheader{Published in \textit{JCAP} \textbf{11} (2026) 005 \quad [\href{https://doi.org/10.1088/1475-7516/2026/11/005}{doi:10.1088/1475-7516/2026/11/005}]}
\makeatother
\title{
Effect of post-recombination accretion on primordial binary black hole mergers within virialized dark matter halos}

\author[a,b]{A. Konovalov,}
\author[b]{K. Postnov}
\affiliation[a]{Faculty of Physics, Lomonosov Moscow State University,
	Leninskie Gory 1-2, Moscow 119991, Russia}
\affiliation[b]{Sternberg Astronomical Institute,  
	Universitetsky pr. 13, Moscow 119234, Russia}

\emailAdd{konovalov.aa22@physics.msu.ru}

\abstract{
Gravitational waves from binary black hole mergers remain among primary sources for ground-based and space laser interferometers. Primordial binary black holes (PBHs) can form binaries in the early Universe or at the matter-dominated stage in virialized dark matter halos at redshifts $z\sim 7-15$. We study enhancement of primordial binary PBH merger rate in virialized dark matter halos due to baryon accretion in the post-recombination epoch within two accretion models onto individual PBH -- Bondi-Hoyle-Littleton (BHL) and Park-Ricotti (PR), representing two extreme cases of the accretion efficiency. We calculate modification of the initial PBH mass function and their fractional contribution to dark matter. A significant enhancement in the binary PBH merger rate is found for the late three-body channel, which was previously treated as subdominant. In the case of delayed virialization onset at $z<10$, merger rates of dynamically formed PBH binaries are limited by the existing LVK binary BH detections. An increased merger frequency is revealed for PBH binaries with highly asymmetric mass ratios. We also calculate expected detection rates of binary PBH mergers for the TianQin space-based gravitational-wave interferometer and show them to be optimistic in the BHL accretion case even for very low initial PBH abundances $f_\text{PBH}\sim 10^{-6}$.}

\begin{document}
\maketitle
\flushbottom

\section{Introduction}
\label{sec:intro}

The hypothesis of primordial black holes (PBHs) formed from cosmological perturbations in the early Universe pioneered in papers \cite{1967SvA....10..602Z,1971MNRAS.152...75H,1974MNRAS.168..399C} became very popular after the detection of the first  binary BH merging event GW150914 by LIGO laser interferometers \cite{2016PhRvL.116f1102A} (see, e.g., Refs. \cite{2016PhRvL.117f1101S,2016PhRvL.116t1301B,2016JCAP...11..036B}). Unlike astrophysical BHs produced in stellar collapses, masses of PBHs can fall within a very broad range from $\sim 10^{15}$~g to supermassive black hole (SMBH) values.
PBHs can form by different mechanisms, including the collapse of primordial density fluctuations during the inflationary epoch, phase transitions at the QCD and electroweak scales, the collapse of cosmic topological defects, scalar field instabilities, etc. \cite{2026arXiv260623846S}.
PBHs can  constitute a fraction of cold dark matter, can seed the growth of SMBH in galaxies and early galaxy formation.  Different PBH formation mechanisms, various astrophysical constraints on the PBH abundance and venues for their detection have been extensively reviewed in the literature \cite{2018PhyU...61..115D, 2021RPPh...84k6902C,2024PhR..1054....1C,2025arXiv250211966G}.

Stellar-mass binary PBHs are especially interesting for the ongoing LIGO/Virgo/KAGRA ground-based gravitational-wave detectors and future space laser interferometers LISA \cite{LISA:2017pwj}, TaiJi \cite{Hu:2017mde}, TianQin \cite{2025CQGra..42q3001L}. 
Although the analysis of the latest GWTC-5 catalog seems to not require 10--50 $M_\odot$ PBHs to explain the observed properties of population of binary BH mergers \cite{2026arXiv260701121L}, binary PBHs with masses $>100\,M_\odot$ can be among primary targets probing fundamental physics by space laser antennas \cite{2026LRR....29....1L}. Therefore, assessing the expected merger rate of such massive binary PBHs is essential. 

Binary PBHs can be formed in the early Universe (early formation channel) or at the matter-dominated  stage due to dynamical interactions (late formation stage) \cite{1998PhRvD..58f3003I,2017PhRvD..96l3523A,2019JCAP...02..018R,2024arXiv240408416R}. Their formation rate as a function of redshift and masses are shaped by their fraction in cold dark matter $f_\text{PBH}$ and mass function $\psi(M)$ either of which are model-dependent. Moreover, the both factors can be modified due to accretion processes at the radiation dominated stage \cite{2020PhRvD.102d3505D}, which in turn are also model-dependent \cite{2025JCAP...08..006J}. 

In this paper, we will focus on the initially log-normal PBH mass function that follows from the baryonic charge fluctuations in the modified Affleck-Dyne baryogenesis \cite{1993PhRvD..47.4244D,2009NuPhB.807..229D}. 
In addition to proving a physical mechanism leading to the log-normal mass function, this model of inhomogeneous baryogenesis addresses simultaneously several other cosmological and astrophysical issues, including the appearance of massive PBHs with $M$ up to $10^4\,M_\odot$ that may seed early SMBH growth \cite{2016JCAP...11..036B}, the existence and appearance of antistars in the Universe and in the Galaxy \cite{2015PhRvD..92b3516B,2023JCAP...08..027B}. The effect of the log-normal PBH mass function on the merger rate of intermediate-mass binary PBHs by space laser interferometers was studied in \cite{2024arXiv240716373P}.

\added{Accretion onto stellar-mass PBHs should affect ionization and thermal history of the Universe after recombination \cite{2008ApJ...680..829R,2017PhRvD..95d3534A,2020PhRvD.102d3505D,2025A&A...699A..49C}. The accretion constraints on the PBH abundance strongly depend on the accretion physics adopted and possible clustering of the PBH population \cite{2017PhRvD..96h3524P}. The initial PBH clustering \textit{per se} affects mergings of binary black holes \cite{2024PhRvD.109l3546S}. In our study we will ignore this effect due to a small PBH abundance assumed $f_\text{PBH}\ll 1$.}

The baryonic accretion onto PBHs from the external environment after recombination can be carried out according to two main scenarios: the inflow of matter in binary systems affecting their orbital decay \cite{2017PhRvD..96l3523A}, and accretion onto isolated PBHs that later form binary PBHs 
in virialized dark matter halos (VDMHs) at later stages of the Universe evolution. For the latter scenario, a more rigorous description can be provided through analytical approximations.

As a population phenomenon, the PBH merger process is directly dictated by the underlying mass function. In the state of art literature on PBH mergers, the mass function is conventionally treated as constant, reflecting the initial conditions at formation. In this work, the post-recombination accretion onto isolated PBHs is coupled to their subsequent mergers in VDMHs, based on a comparative evaluation of the Bondi-Hoyle-Littleton (BHL) and Park-Ricotti (PR) 
frameworks reviewed in \cite{2025JCAP...08..006J}. These two accretion models can be considered as limiting cases, with BHL being the most effective and PR the least effective one. 

Our aim is to quantify the enhancement in the merger rate of the binary PBHs formed in the late channel due to post-recombination baryon accretion and determine whether this mechanism can raise the analyzed channel from a subdominant status (without accretion, \cite{2024arXiv240408416R}) to potentially detectable by the TianQin space laser interferometer. We find that the BHL accretion by significantly modifying the heavy tail of the PBH mass distribution enables a strong enhancement in the late dynamical formation of binary PBHs. In the BHL accretion mode, the current LVK estimates of binary black holes merger rate set upper limits on the PBH abundance $f_\text{PBH}$ and give optimistic prospects of detections by space laser interferometers like TianQin even for small initial PBH abundance $f_\text{PBH}\sim 10^{-6}$.

\section{Accretion models}
\label{sec:accretion}

This Section reviews current models of baryon matter accretion onto PBHs: the Bondi--Hoyle--Littleton model, which accounts for cosmological factors, and the Park--Ricotti model, which takes into account local physical processes around accreting PBHs. These accretion models are analyzed for the epoch spanning from post-recombination up to the moment of VDMHs formation at $z\sim 7-15$.

\subsection{Separability of accretion and mergers}
\label{sec:accretion:separate}
Earlier stages of the cosmic evolution are excluded from consideration because the matter existed in the state of a relativistic plasma. In this regime, the sound speed \added{$c_s = \frac{1}{\sqrt3} c$} is comparable to the speed of light, which suppresses significant accretion, $\dot{M} \propto \frac{G^2 M^2}{c_s^3}$~\cite{2025JCAP...08..006J}. Therefore, the recombination redshift $z_\text{rec} \sim 1100$ is adopted as the starting point for numerical calculations.

Additionally, the cut-off point of efficient accretion onto isolated PBHs must be specified. As the redshift decreases, large-scale cosmological structures begin to form. Subsequently, dark matter halos decouple from the Hubble flow and start virializing. The velocity dispersions of PBHs within these halos significantly exceed the relative velocities between the black holes and gas prior to their infall into the halos~\cite{2025JCAP...08..006J}. This kinematic feature allows the contribution of accretion within VDMHs to be neglected compared to the preceding stage (the dependence of the accretion rate on the relative velocity for both models will be demonstrated below in this Section).

Since this study considers double PBH formation and mergers exclusively within the late-time channel (i.e., after their infall into VDMHs), the stage of accretion mass growth and the subsequent mergers of dynamically formed BH binaries can be solved independently. Accretion is calculated prior to virialization, whereas mergers are evaluated using the PBH population parameters modified by the preceding accretion process. The onset of VDMH formation is treated as a free parameter; by analogy with~\cite{2020PhRvD.102d3505D}, the corresponding redshifts $z_\text{cut-off} = \{7, 10, 15\}$ are considered, where $z_\text{cut-off}=15$ corresponds to a conservative estimate and $z_\text{cut-off}=7$ to an optimistic one. As a simplification, the halo virialization is assumed to be an instantaneous process (this approximation will be discussed in Section~\ref{sec:discussion}), which allows the PBH population characteristics to remain constant during the merger analysis.

\subsection{BHL accretion}
\label{sec:accretion:BHL}

The Bondi-Hoyle-Littleton (BHL) model is a classical accretion approximation based on the assumptions of negligible gas self-gravitation and invariant gas parameters specified by the boundary conditions at infinity and ignores radiative feedback. The model considers axially symmetric accretion of gas onto an isolated BH of mass $M$, with the rate given by:
\begin{equation}
\dot{M}_\text{BHL} = 4\pi \lambda \rho_B r_B^2 v_\text{eff}
\end{equation}
where $v_\text{eff} = \sqrt{v_\text{PBH}^2+c_s^2}$ is the effective velocity incorporating the sound speed $c_s$ in the gas and the relative velocity of the BH $v_\text{PBH}$, $r_B = \frac{G M}{v_\text{eff}^2}$ is the Bondi radius defining the effective gravitational capture boundary for baryons, and the baryon matter density is given by $\rho_B = \Omega_B \rho_c (1+z)^3$ as a fraction $\Omega_B$ of the critical density of the Universe $\rho_c = \frac{3H_0^2}{8\pi G}$.

In modern studies, the parameter $\lambda$, which denotes the accretion efficiency, is no longer considered as a free parameter. A modification of this parameter was proposed by Ricotti et al.~\cite{2008ApJ...680..829R} to account for cosmological effects that suppress the mass accretion rate: the impossibility of gas capture from regions extending beyond the Hubble horizon and Compton drag induced by the cosmic microwave background (CMB):
\begin{equation}
\lambda = \exp\left(\frac{9/2}{3+\beta^{3/4}}\right)\,x_\text{cr}^2,
\end{equation}
where the dimensionless critical radius is introduced\footnote{An equivalent formula, $x_{\text{cr}} = \frac{-1+\sqrt{1+\beta}}{\beta}$, is frequently encountered in the literature; however, it is less convenient for numerical simulations due to the $0/0$ indeterminacy as $\beta \to 0$.}:
\begin{equation}
	x_\text{cr} = \frac{r_\text{cr}}{r_B} = \frac{1}{1+\sqrt{1+\beta}}
\end{equation}
and the dimensionless parameter that effectively accounts for the gas viscosity given by:
\begin{equation}
\beta = \frac{M}{10^4\, M_\odot} \left(\frac{1+z}{1000}\right)^{3/2} \left(\frac{v_\text{eff}}{5.74\,\text{km}\,\text{s}^{-1}}\right)^{-3} \left[0.257+1.45\frac{x_e}{0.01}\left(\frac{1+z}{1000}\right)^{5/2}\right]
\end{equation}

It can be shown that as $\beta \to \infty$, the accretion efficiency exhibits the asymptotic behavior $\lambda\big|_{\beta \to \infty} = \frac{1}{\beta}$, which, due to the structure of the equation for $\beta$, leads to an inverse proportionality with respect to the PBH mass $M$. Consequently, a transition occurs from the quadratic Bondi regime to a linear regime analogous to the Eddington limit: $\dot{M}\big|_{M \to \infty} \propto \lambda\big|_{M \to \infty} \times M^2 \propto M$.

The profiles of the sound speed $c_s$, PBH velocity $v_\text{PBH}$, and ionization fraction $x_e$ are chosen in accordance with Ricotti et al.~\cite{2008ApJ...680..829R} (ROM8). Note, alternative velocity profiles are used in several studies of the post-recombination accretion (see Section~\ref{sec:discussion} for details).

%Since this study analyzes the evolution on the redshift scale, it is necessary to 
Recast the time-evolution equations in terms of redshift using the relation for the $\Lambda\text{CDM}$ model of the Universe dominated by non-relativistic matter and dark energy:
\begin{equation}
	\dot{z} = - H_0 (1+z) \sqrt{\Omega_M(1+z)^3 + \Omega_\Lambda}
\end{equation}

The cosmological parameters are set to the Planck 2018~\cite{2020A&A...641A...6P} values: the dark energy density parameter $\Omega_\Lambda = 0.685$, the non-relativistic matter density parameter $\Omega_M = 0.315$, the baryon matter density parameter $\Omega_B = 0.0493$, and the Hubble constant $H_0 = 67.4\,\text{km}\,\text{s}^{-1}\,\text{Mpc}^{-1}$.

\subsection{PR accretion}
\label{sec:accretion:PR}

In contrast to the BHL model, the Park--Ricotti (PR) model~\cite{2020MNRAS.495.2966S} is governed by the gas parameters in the local environment surrounding the BH, which are modified by radiative feedback (predominantly X-ray and UV radiation), rather than global cosmological factors. According to numerical simulations, BHs can be surrounded by an ionization bubble that severely restricts the inflow of additional matter.

The ionization front divides the space into two regions: internal and external. In the external region, the gas parameters remain unmodified and correspond to the mean cosmological values at the current epoch. In the internal region, the gas is heated due to radiation feedback, which leads to changes in the local sound speed, density, and, consequently, the relative velocity of the PBH. The accretion rate in the PR model can be expressed in the BHL formalism using local parameters:
\begin{equation}
	\dot{M}_\text{PR} = 4\pi \rho_B^\text{in} r_B^2(v_\text{eff}^\text{in}) v_\text{eff}^\text{in},
\end{equation}
where the superscript 'in' denotes the values of the quantities inside the ionization zone.

In this model, the accretion rate is no longer monotonic with respect to the BH velocity; instead, the profile exhibits a prominent peak (or cusp) that reaches its maximum at $v_\text{PBH} = v_R$. In the BH rest frame, the gas velocity is represented by a piecewise function:
\begin{equation}
	v_\text{PBH}^\text{in} = \begin{cases}
		\frac{\rho_B}{\rho_B^\text{in}} v_\text{PBH} & \text{for } v_{\text{PBH}} \leq v_D \, ,\\
		c_s^\text{in} & \text{for } v_D < v_\text{PBH} < v_R \, , \\
		\frac{\rho_B}{\rho_B^{\text{in}}} v_\text{PBH} & \text{for } v_R \leq  v_\text{PBH} \,.
	\end{cases}
\end{equation}

The boundary velocities, $v_D = c_s^\text{in} - \sqrt{c_s^\text{in 2} - c_s^2}$ and $v_R = c_s^\text{in} + \sqrt{c_s^\text{in 2} - c_s^2}$, correspond to two distinct ionization front types: D-type and R-type. Under D-type front conditions, the gas has sufficient time to dynamically respond to the heating, whereas under R-type conditions it does not, rendering the latter reminiscent of the classical Bondi regime; in the intermediate case, a self-regulated accretion regime is established. These boundary velocities are determined as the roots of the following equation:
\begin{equation}
	\Delta \equiv (c_s^2+v_\text{R/D}^2)^2 - 4 \, c_{\text{s}}^{\text{in 2}} v_\text{R/D}^2 = 0
\end{equation}

The gas density inside the ionization front is also given by a piecewise function:
\begin{equation}
	\rho_B^{\text{in}} = \begin{cases}
		\rho_B \left( \frac{v_\text{PBH}^2 + c_s^2 + \sqrt{\Delta}}{2 c_s^\text{in 2}} \right) & \text{for } v_\text{PBH} \leq v_D \, ,\\
		\rho_B \left( \frac{v_\text{PBH}^2 + c_s^2}{2 c_s^\text{in 2}} \right) & \text{for } v_D < v_\text{PBH} < v_R \, , \\
		\rho_B \left( \frac{v_\text{PBH}^2 + c_s^2 - \sqrt{\Delta}}{2 c_s^\text{in 2}} \right) & \text{for } v_R \leq  v_\text{PBH} \,.
	\end{cases}
\end{equation}

The primary uncertainty of this model lies in the choice of the sound speed contrast constant between the internal and external regions. By analogy with~\cite{2021MNRAS.505.4036S}, this study adopts $c_s^\text{in} = 25 c_s$; however, the specific numerical value of this constant can significantly influence the accretion rate (see Section~\ref{sec:discussion} for details).

Under the condition of a substantial thermal contrast, $\frac{c_s^\text{in}}{c_s} \gg 1$, the expressions for the boundary velocities can be approximated by simpler relations: $v_D = \frac{c_s^2}{2 c_s^\text{in}}$ and $v_R = 2 c_s^\text{in}$. It is worth noting that for the adopted ROM8 velocity profile, the relative velocity of the BH outside the ionization front always remains within these boundary limits. Nevertheless, regimes with extremely high and low velocities must be taken into account during the luminosity averaging, which is described in detail in Section~\ref{sec:accretion:factors} below.

\subsection{Factors enhancing accretion}
\label{sec:accretion:factors}

This Subsection describes additional factors that are excluded from the classical formulations of both the BHL and PR models but are intrinsically common to both: the surrounding DM mini-halos of the PBHs and statistical velocity averaging. Accounting for these factors can lead to a substantial enhancement in the PBH accretion rate.

\begin{itemize}
    \item According to~\cite{2026NCimR.tmp....4C}, the PBH mass fraction in DM must be significantly less than unity across a wide mass range. Consequently, dark matter (DM) mini-halos are expected to form around a BH of mass $M$~\cite{2008ApJ...680..829R}, thereby deepening the gravitational potential of the accreting BH to an effective value of $M_\text{eff} = M + M_h$. The mass of the mini-halo can be represented as:
    \begin{equation}
	   M_h = 3M \left( \frac{1+z}{1000} \right)^{-1}.
    \end{equation}
    Minor corrections associated with halo smoothing are also utilized (assuming a DM density profile of $\rho \sim r^{-\alpha}$ with $\alpha = 9/4$).
    \item The accretion rate and hence the BH luminosity strongly depend on the relative velocity. In the general case, this dependence can be represented as a function of the effective velocity, $g(v_\text{eff})$. Since the relative velocity of the PBH is not constant but follows a Maxwellian distribution $f_M(v, \sigma)$ with a dispersion $\sigma = \langle v_\text{rel} \rangle$, a luminosity-weighted speed is introduced~\cite{2008ApJ...680..829R}:
    \begin{equation}
    	\langle g (v_\text{eff})\rangle \equiv \int_0^\infty g(v_\text{eff}(v))f_M(v, \sigma)\,dv.
    \end{equation}
    
    For the BHL model, an analytical formula for the effective velocity can be derived depending on the accretion regime (the presence or absence of a disk structure):
    \begin{equation}
    	v_\text{eff} = c_s \begin{cases}
    		2^{2/3} \left[\frac{\sqrt{2 \pi } e^{\frac{1}{2 \mathcal{M}^2}} \left(\mathcal{M}^4-2 \mathcal{M}^2-1\right) \text{erfc}\left(\frac{1}{\sqrt{2} \mathcal{M}}\right)+2 \mathcal{M}\left(\mathcal{M}^2+1\right)}{\mathcal{M}^7}\right]^{-1/6} & \text{for } \dot{m} \leq 1\, , \\
    		2^{1/2}\left[\frac{e^{\frac{1}{4 \mathcal{M}^2}} \left(\left(2 \mathcal{M}^2+1\right) K_0\left(\frac{1}{4 \mathcal{M}^2}\right)-K_1\left(\frac{1}{4 \mathcal{M}^2}\right)\right)}{\sqrt{\pi }\mathcal{M}^5}\right]^{-1/3} & \text{for } \dot{m} > 1 \, ,
    	\end{cases}
    \end{equation}
    where $\dot{m} = \frac{\dot{M}}{\dot{M}_\text{Edd}}$ is the accretion rate normalized to the Eddington rate, $\mathcal{M} = \frac{\sigma}{c_s}$ is the Mach number, $K_0$ and $K_1$ are the modified Bessel functions of the second kind of zeroth and first order, respectively, and $\text{erfc}(x) = 1 - \text{erf}(x)$ is the complementary error function.
    
    For the PR model, obtaining an analytical solution is not feasible; therefore, the averaging is performed numerically.
\end{itemize}

\section{Evolution of population parameters}
\label{sec:population}

In this section, PBHs are treated as a population as a whole rather than as isolated objects. In this context, we investigate the evolution of the population parameters due to the preceding post-recombination accretion process. Specifically, we focus on the reshaping of the mass function, as well as the evolution of the mean PBH mass and the PBH fraction in DM.

In this work, we consider\footnote{This specific parametrization traces back to the Dolgov-Silk, $\frac{dN}{dM} = \mu^2\exp\left[-\gamma\log^2\frac{M}{M_0}\right]$~\cite{1993PhRvD..47.4244D}, where $M_0 = \langle M\rangle_i \exp\left[-\frac{3}{4\gamma}\right]$ is the peak mass. In the literature, an alternative parametrization in terms of the width parameter $\sigma$ is also employed, which is uniquely related to $\gamma$ via $\gamma = 1 / (2 \sigma^2)$.} the initial log-normal mass function
\added{\begin{equation}
	\psi_i(M) = \frac{1}{M}\sqrt{\frac{\gamma}{\pi}}\exp{\left[-\gamma\left(\log{\frac{M}{\langle M\rangle_i}}+\frac{1}{4\gamma}\right)^2\right]},
\end{equation}}
normalized to mass, $\int \psi(M) \,dM = 1$. The parameters are expected to lie within the ranges $\langle M\rangle_i = 10-30\, M_\odot$~\cite{2020JCAP...07..063D} and $\gamma = 0.6-1.5$\,\cite{2016JCAP...11..036B}, where $\langle M\rangle_i$ characterizes the mean mass of the distribution, and $\gamma$ determines its sharpness. Hereafter, the subscript 'i' denotes values referring to the epoch of recombination, while the subscript 'z' utilized later refers to an arbitrary redshift.

As shown in Ref.~\cite{2016JCAP...11..036B}, PBHs can form with masses up to $10^4\,M_\odot$. For numerical calculations, we adopt the initial mass range $\log_{10} (M_i/M_\odot) \in [-3, 4]$. Due to the exponential decay of tails of the distribution, this choice of the left mass bound ensures a normalization that is strictly close to unity.

\subsection{Reshaping of the mass functions}
\label{sec:population:massfunc}

By the onset of virialization, the PBH masses grow as $m_i \rightarrow m(m_i, z_\text{cut-off})$. According to the probability conservation law, their distribution evolves simultaneously as well:
\begin{equation}
	\psi(m) = \psi_i(m_i) \left.\frac{\partial m_i}{\partial m} \right|_{z_\text{cut-off}},
\end{equation}
where the Jacobian accounts for the mapping between the initial and evolved masses at the cut-off redshift.

\begin{figure}[htbp]
	\centering
	\includegraphics[width=.45\textwidth]{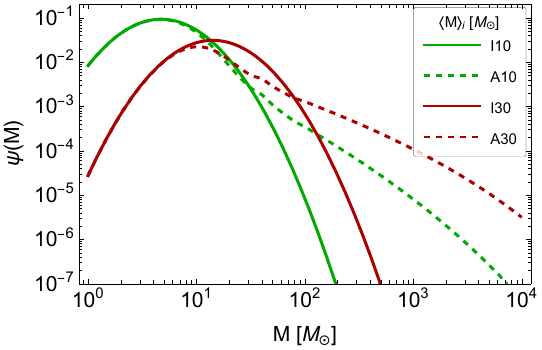}
	\qquad
	\includegraphics[width=.45\textwidth]{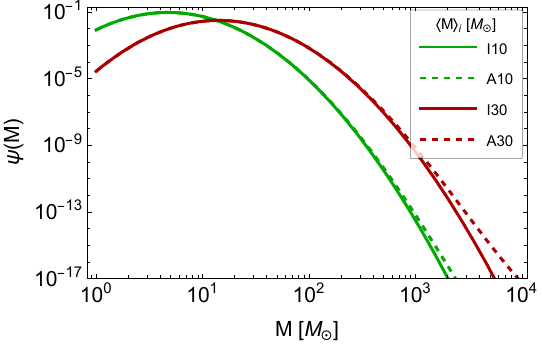}
	\caption{Reshaping of the mass function distribution $\psi(m)$ at fixed $z_\text{cut-off} = 10$ and $\gamma = 1$ due to post-recombination accretion for the BHL (left panel) and PR (right panel) models. Solid lines correspond to the initial ('I', log-normal) distributions, dashed lines to the evolved ('A') ones; green lines correspond to $\langle M\rangle_i = 10\, M_\odot$ and maroon lines to $\langle M\rangle_i = 30\, M_\odot$.
\label{sec:population:massfunc:fig_reshape1}}
\end{figure}

Analysis of the resulting distributions (see Figure~\ref{sec:population:massfunc:fig_reshape1}) shows that for certain parameters (in particular, for sufficiently wide initial distributions with $\gamma<1$), the heavy-mass tail can be noticeably enhanced. Furthermore, within the BHL model at high initial mean masses, the high-mass tail of the spectrum can reshape into a plateau spanning a wide mass range starting from $\sim 100\, M_\odot$.

The dynamical binary PBH formation and differential merger rate in VDMH (see below in Section~\ref{sec:mergerrate}) depend on the accretion-modified PBH mass distributions $\psi(m_1)\psi(m_2)$, the total mass of the binary $M=m_1+m_2$, and the symmetric mass ratio, $\eta=\frac{m_1m_2}{(m_1+m_2)^2}$,  $\eta \in (0, 1/4]$, where the maximum is reached for equal-mass components. In the case of highly mass asymmetry, $\eta \to q = m_1/m_2 \ll 1$. Therefore, it is useful to calculate the binary mass function $\psi(M)\psi(\eta)$ that strongly affects the differential binary PBH merger rate.

%The binary mass function $\psi(M)\psi(\eta)$ is also considered, which, as will be shown later in Section~\ref{sec:mergerrate}, enters as a factor into the differential merger rates of binary PBHs. This function represents the distribution over the total mass $M = m_1 + m_2$ and the symmetric mass ratio

% $\eta = \frac{m_1m_2}{(m_1+m_2)^2}$.

It can be derived by changing variables $(m_1, \, m_2) \rightarrow (M,\eta)$:
\begin{equation}
    \label{eq:Jac}
    \psi(M)\psi(\eta) = \left| \frac{\partial(m_1, m_2)}{\partial(M, \eta)} \right| \psi(m_1)\psi(m_2)
\end{equation}
where the Jacobian $\left|\frac{\partial(m_1, m_2)}{\partial(M, \eta)}\right| = \frac{M}{\sqrt{1-4\eta}}$ accounts for the mapping between the evolved masses and the binary system parameters at the cut-off redshift.

\begin{figure}[htbp]
	\centering
	\includegraphics[width=.45\textwidth]{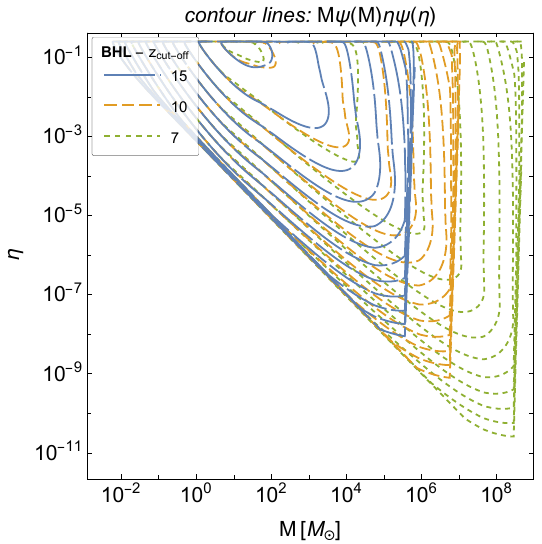}
	\qquad
	\includegraphics[width=.45\textwidth]{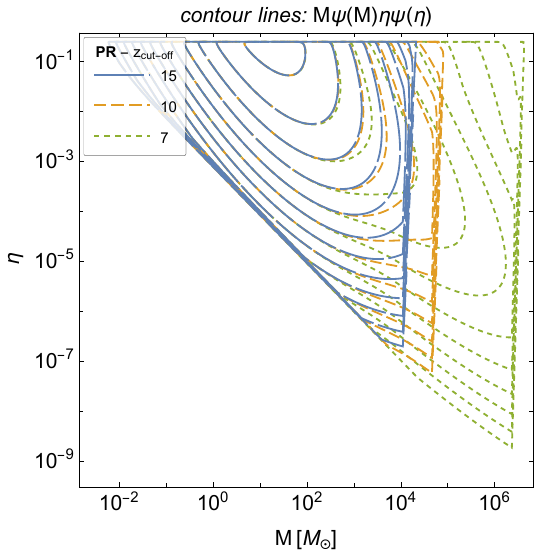}
	\caption{Reshaping of the binary mass function $\psi(M)\psi(\eta)$ at fixed  $\langle M\rangle_i = 20\, M_\odot$ and $\gamma = 1$ due to post-recombination accretion. The left and right panels correspond to the BHL and PR accretion models, respectively.\label{sec:population:massfunc:fig_reshape2}}
\end{figure}

Analysis of the obtained binary mass functions (see contour maps in Figure~\ref{sec:population:massfunc:fig_reshape2}) points to two processes (expressed as shifts along two axes) as the accretion time available for the PBH mass growth increases \mbox{$(z_\text{cut-off}: 15 \rightarrow 7)$}:
\begin{itemize}
    \item shift along the total mass axis $M$, caused by population-wide mass growth due to post-recombination accretion, which leads to the interaction of heavier components;
    \item shift along the symmetric mass ratio axis $\eta$, induced by the non-linear nature of accretion (negligible growth at low and significant growth at high masses), which results in a flattening of the mass function and increases the probability of interactions involving asymmetric components.
\end{itemize}

\subsection{Growth of the PBH fraction in DM}
\label{sec:population:fraction}

Driven by the evolution of the spectral shape, the mean mass of the population increases. For the adopted normalization, the mean mass $\langle M \rangle$ at a given redshift $z$ is calculated as follows:
\begin{equation}
	\langle M \rangle_z = \int  m(m_i,\, z) \, \psi_i(m_i)\, dm_i
\end{equation}
where $m(m_i,\, z)$ is the solution to the accretion equation for an isolated BH with an initial mass $m_i$. Furthermore, the integration variable in this expression is transformed using the Jacobian discussed above. This approach improves the computational accuracy by eliminating the need to numerically evaluate the derivative $\frac{\partial m_i}{\partial m}$.

The evolution of the mean mass over the distribution is directly coupled to the evolution of the PBH (initially unassociated) mass fraction in DM $f_\text{PBH}$, which can be defined as follows\footnote{\added{Here and below the 'PBH' subscript for the PBH mass fraction in DM is omitted for readability.}}:
\begin{equation}
	f_\text{z} = \frac{\Omega_\text{PBH z}}{\Omega_\text{DM z}} = \frac{M_\text{PBH i}+M_\text{acc z}}{M_\text{DM i}+M_\text{acc z}} = \frac{f_\text{i} \frac{\langle M\rangle_z}{\langle M\rangle_i}}{1+f_\text{i}\left(\frac{\langle M\rangle_z}{\langle M\rangle_i} - 1\right)}
\end{equation}
where $M_\text{PBH i}$ and $M_\text{DM i}$ are the total masses of PBHs and DM at the recombination epoch, respectively, and $M_\text{acc z}$ is the accumulated accretion mass of PBHs due to post-recombination accretion up to the redshift $z$. Furthermore, factors of $(1+z)$ cancel out in this expression because both the PBH number density (inverse volume) and the matter density contain the same scale factor dependence $(1+z)^3$.

According to current observational constraints~\cite{2026NCimR.tmp....4C}, the fraction of stellar-mass PBHs in DM must be significantly less than unity, $f_z \ll 1$, which consequently implies $f_i < f_z$. Under these conditions, the depletion of baryon matter can be neglected, thereby reducing the growth to the linear regime: %Carr cite
\begin{equation}
	f_z \Big|_{f_i \to 0}  = f_i \frac{\langle M\rangle_z}{\langle M\rangle_i}
\end{equation}

\begin{figure}[htbp]
	\centering
	\includegraphics[width=.45\textwidth]{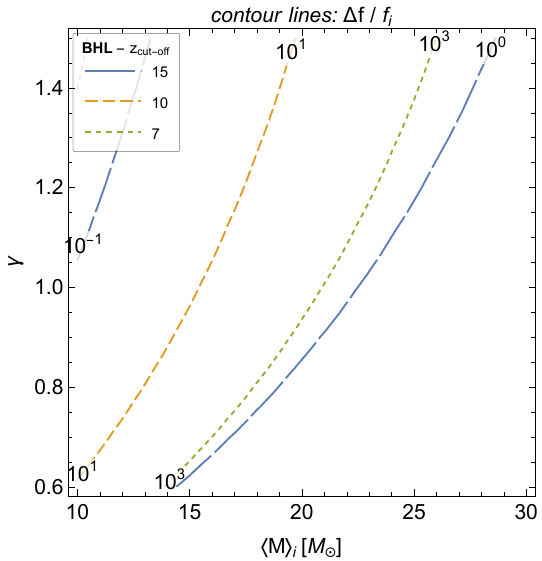}
	\qquad
	\includegraphics[width=.45\textwidth]{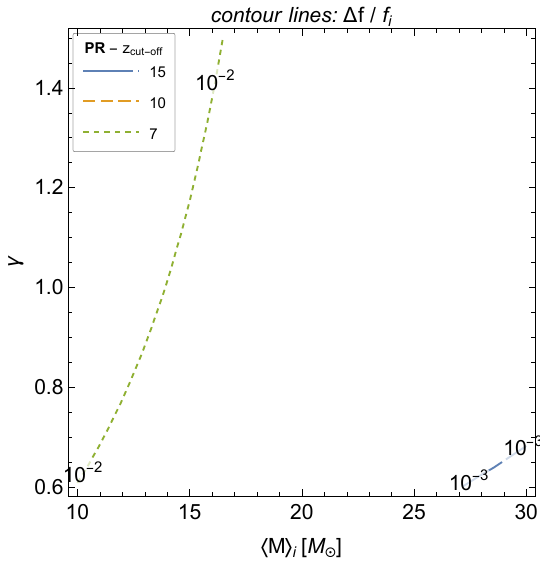}
	\caption{Enhancement of the initial PBHs fraction in DM $\frac{\Delta f}{f_i}$ due to post-recombination accretion as a function of the initial distribution parameters $\langle M\rangle_i,$ and $\gamma$. The left and right panels correspond to the BHL and PR accretion models, respectively.\label{sec:population:fraction:fig}}
\end{figure}

As demonstrated by the calculated dependencies of $\frac{\Delta f}{f_i} = \frac{f_z}{f_i} -1$ plotted in Figure~\ref{sec:population:fraction:fig}, the general trend is an increase in the PBH mass fraction in DM with decreasing sharpness $\gamma$ of the initial distribution and increasing initial mean mass $\langle M \rangle_i$. It is worth noting that in the BHL model, this growth can be significant (by several orders of magnitude, depending on the virialization onset redshift $z_\text{cut-off}$), whereas in the PR model the growth does not exceed a few percent.
\section{Merger of binary PBHs in virialized DM halos}
\label{sec:mergerrate}

This Section describes the model of the dynamic formation of PBH binaries in VDMHs (the late-time channel) and presents the results of numerical integration of PBH merger rates for the considered initial distribution parameters.

For numerical calculations, fiducial initial values  $f_\text{ref} = 10^{-6}$ and $10^{-3}$ are fixed for the BHL and PR accretion models, respectively. As shown in Section~\ref{sec:population:fraction}, the PBH mass fraction in DM scales linearly; thus, all plots presented in this and the subsequent Section~\ref{sec:tianqin} can be scaled to the desired $f_i$ by multiplying by the factor $\left(\frac{f_i}{f_\text{ref}}\right)^n$. Here, $n = 2$ or $3$ is used for the two- or three-body PBH binary formation channel, respectively (see below in this Section for details).

\subsection{Enhancement of merger rates}
\label{sec:mergerrate:enh}

This study utilizes the analytical expressions derived by Raidal et al.~\cite{2024arXiv240408416R}. The authors identify four PBH merger channels based on the time of binary formation (early- and late-time channels) and by the number of interacting bodies (two- and three-body channels). The early-time channel describes PBHs that formed bound systems immediately after their formation; this channel is found to be dominant. In contrast, the late-time channel characterizes PBHs that form binaries within VDMHs; this channel is considered subdominant. The binary formation mechanism in the two-body channel is a close encounter between two PBHs, accompanied by the dissipation of excess energy via gravitational wave (GW) emission. In the three-body channel, the mechanism involves a close encounter of three PBHs, where the excess energy is transferred to one of the PBHs, which is subsequently gravitationally ejected from the system.

The differential PBH merger rates (per unit square mass of the merging components) in the late two- and three-body channels, respectively read: 
%by the following analytical formulas:
\begin{align}
    \label{R2}
    \frac{d \mathcal{R}_{L2}}{dm_1 dm_2} & = \frac{3.4\times 10^{-6}}{\text{Gpc}^3\,\text{yr}}f_\text{PBH}^2 \delta_\text{eff}\left(\frac{\sigma_v}{\text{km} \,\text{s}^{-1}}\right)^{-11/7} \eta^{-5/7}\psi(m_1)\psi(m_2)\,, 
    \\[1ex]
    \label{R3}
    \frac{d \mathcal{R}_{L3}}{dm_1 dm_2} &= \frac{1.3\times 10^{-16}e^{-6(\gamma_a-1)}}{\text{Gpc}^3\,\text{yr}}f_\text{PBH}^3 \delta_\text{eff}^2\left(\frac{\sigma_v}{\text{km} \,\text{s}^{-1}}\right)^{-9+\frac{8\gamma_a}{7}} \nonumber 
    \\ 
    &\qquad \times \eta^{-1+\frac{\gamma_a}{7}}\left(\frac{M}{M_\odot}\right)^{3-\frac{\gamma_a}{7}}\left(\frac{t}{t_{z=0}}\right)^\frac{\gamma_a}{7}\mathcal{F}\left(\frac{\langle M \rangle}{2 \eta M}\kappa_\text{min}\right)\psi(m_1)\psi(m_2)
\end{align}
where the function $\mathcal{F}$ with the hardness parameter $\kappa_\text{min} = 5$:
\begin{equation}
	\mathcal{F}(\kappa) = \kappa^{-4+\frac{4\gamma_a}{7}} \left[ \frac{6\sqrt{\pi}}{7-\gamma_a} + \frac{72}{63-8\gamma_a}\kappa^{-\frac{1}{2}} + \frac{15\sqrt{\pi}}{70-8\gamma_a}\kappa^{-1} + \frac{22}{77-8\gamma_a}\kappa^{-\frac{3}{2}} \right]
\end{equation}
and the cosmological time is
\begin{equation}
	t(z) = \frac{2}{3 H_0 \sqrt{\Omega_\Lambda}} \ln \left[ \frac{\sqrt{\Omega_\Lambda} + \sqrt{\Omega_M (1+z)^3 + \Omega_\Lambda}}{\sqrt{\Omega_M}} \right]
\end{equation}

Following Raidal et al.~\cite{2024arXiv240408416R}, the halo parameters in this study are fixed as follows: the local density contrast $\delta_\text{eff} = 10^5$ and $10^7$ for the two- and three-body channels, respectively, the velocity dispersion $\sigma_v = 1\,\text{km} \,\text{s}^{-1}$, and the angular momentum distribution parameter $\gamma_a = 2$, which corresponds to a thermal distribution.

As a result of preceding accretion, both the PBH mass fraction in DM $f_\text{PBH}$ and the mean mass over the distribution $\langle M \rangle$ increase by the redshift $z_\text{cut-off}$; furthermore, the mass spectrum $\psi(M)$ itself is reshaped. Since the integrated merger rate is a multi-parametric function $\mathcal{R} = \mathcal{R}(\langle M \rangle,\, \gamma,\,f_0,\, z_\text{cut-off})$, a merger rate enhancement factor $\mathcal{E} $ is introduced, defined as the ratio of the merger rates with and without accounting for prior accretion:
\begin{equation}
	\mathcal{E}(z) = \frac{\mathcal{R}_\text{acc}(f_i, z)}{\mathcal{R}_\text{w/o acc}(f_i, z)}\bigg|_{f_i \to 0} = \frac{\Delta \mathcal{R}(f_i, z)}{\mathcal{R}_\text{w/o acc}(f_i, z)}\bigg|_{f_i \to 0} + 1
\end{equation}
where the integrated merger rate $\mathcal{R} = \iint d\mathcal{R}(m_1,\,m_2)$. This function is constructed in such a way that it eliminates the parameter of the initial PBH mass fraction in DM $f_i$ as well as the model parameters of VDMHs.

Consequently, the corresponding enhancement factors for the late two- and three-body channels take the form of a product of two factors, where the first represents the increase in the PBH number density in DM $f_\text{PBH}$, and the second accounts for the reshaping of the mass function $\psi(m)$:
\begin{align}
    \label{enh2}
    \mathcal{E}_\text{L2}(z) &= \left(\frac{\langle M \rangle_z}{\langle M\rangle_i}\right)^2\times \frac{\iint dm_1\, dm_2\eta^{-5/7}\psi_z(m_1)\psi_z(m_2)}{\iint dm_1\, dm_2 \eta^{-5/7}\psi_i(m_1)\psi_i(m_2)}\,,
    \\[1ex]
    \label{enh3}
    \mathcal{E}_\text{L3}(z) &= \left(\frac{\langle M \rangle_z}{\langle M\rangle_i}\right)^3\times \frac{\iint dm_1\, dm_2 \eta^{-1+\frac{\gamma_a}{7}}\left(\frac{M}{M_\odot}\right)^{3-\frac{\gamma_a}{7}}\mathcal{F}\left(\frac{\langle M \rangle_z}{2 \eta M}\kappa_\text{min}\right)\psi_z(m_1)\psi_z(m_2)}{\iint dm_1\, dm_2 \eta^{-1+\frac{\gamma_a}{7}}\left(\frac{M}{M_\odot}\right)^{3-\frac{\gamma_a}{7}}\mathcal{F}\left(\frac{\langle M\rangle_i}{2 \eta M}\kappa_\text{min}\right)\psi_i(m_1)\psi_i(m_2)}
\end{align}
where, for readability, the subscripts 'i' and 'z' corresponding to the initial and arbitrary redshifts are omitted for $m_1$, $m_2$, and their dependent functions.

\begin{figure}[htbp]
	\centering
	\includegraphics[width=.45\textwidth]{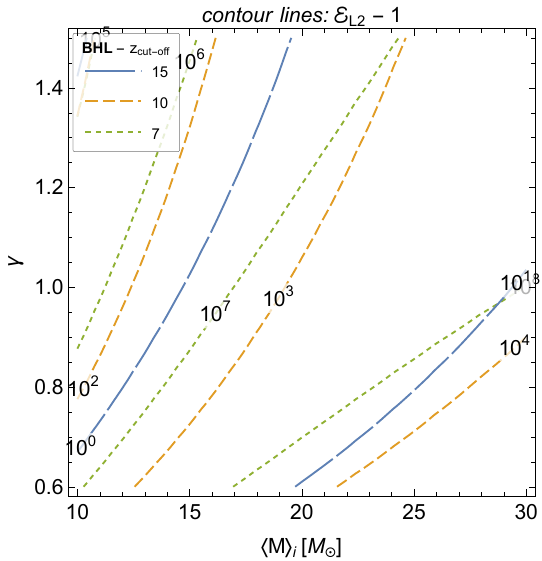}
	\qquad
	\includegraphics[width=.45\textwidth]{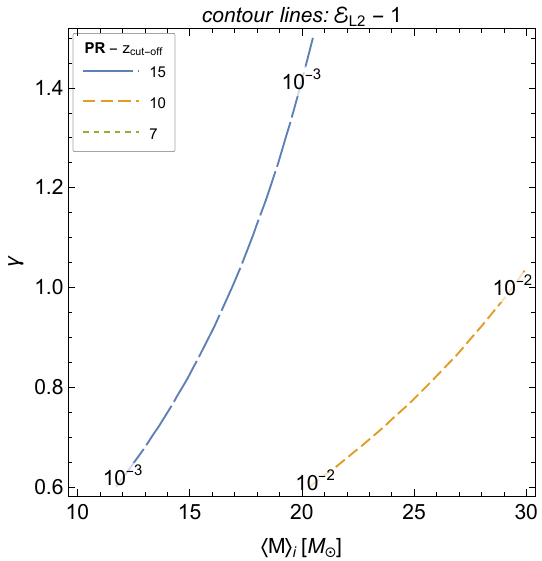}
	
	\includegraphics[width=.45\textwidth]{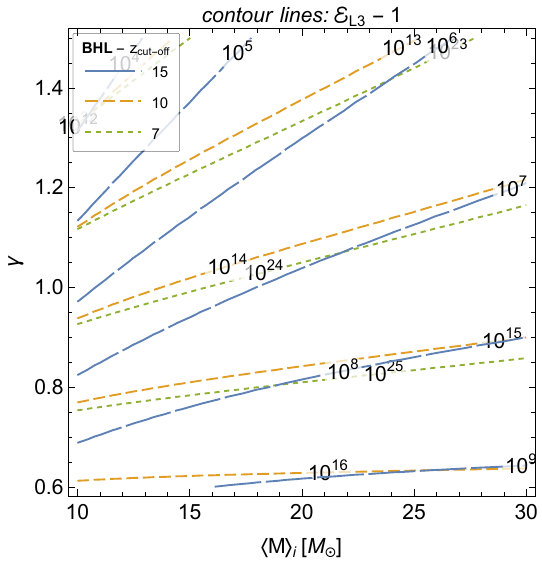}
	\qquad
	\includegraphics[width=.45\textwidth]{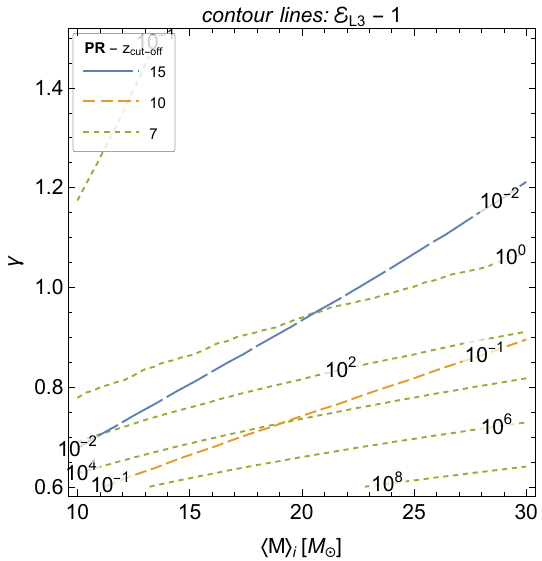}
	\caption{The gain coefficient $\mathcal{E} - 1$ of the merger rate of binary PBHs as a function of the initial distribution parameters $\langle M\rangle_i,$ and $\gamma$. Rows correspond to the two-body (top) and three-body (bottom) channels, while columns represent the BHL (left) and PR (right) accretion models.\label{sec:mergerrate:enh:fig}}
\end{figure}

Analysis of the obtained dependencies in Figure~\ref{sec:mergerrate:enh:fig} indicates a moderate increase of the enhancement factor in the two-body channel and a substantial growth in the three-body channel.
%\added{(which strongly depends on the total mass $M$))}. 
This behavior is consistent with the structure (positive powers of mean $\langle M\rangle$ and total $M$ masses and negative powers of symmetric mass ratio $\eta$) of the corresponding equations (\ref{enh2},~\ref{enh3}) for the enhancement factors $\mathcal{E}$ in the late two- and three-body channels, respectively:
\begin{align*}
    \frac{d\mathcal{R}_\text{L2 acc}}{dm_1 dm_2}\bigg/\frac{d\mathcal{R}_\text{L2 w/o acc}}{dm_1 dm_2} & \propto \left(\frac{\langle M \rangle_z}{\langle M\rangle_i}\right)^2 \times \left(\frac{\eta_z}{\eta_i}\right)^{-5/7}\,,
    \\[1ex]
    \frac{d\mathcal{R}_\text{L3 acc}}{dm_1 dm_2}\bigg/\frac{d\mathcal{R}_\text{L3 w/o acc}}{dm_1 dm_2} & \propto \left(\frac{\langle M \rangle_z}{\langle M\rangle_i}\right)^3 \times \left(\frac{\eta_z}{\eta_i}\right)^{-1+\frac{\gamma_a}{7}} \left(\frac{M_z}{M_i}\right)^{3-\frac{\gamma_a}{7}}\,,
\end{align*}
where $\gamma_a/7 < 1$ for the fixed angular momentum distribution parameter $\gamma_a = 2$ as well as within its extended range (see Section~\ref{sec:discussion}). Furthermore, the accretion end redshift $z_\text{cut-off}$ much more strongly affect the results than the parameters $\langle M\rangle_i, \, \gamma$ of the initial mass function.

\subsection{Reshaping of differential merger rates}
\label{sec:mergerrate:difrate}

The differential rates, modified by post-recombination accretion, are also considered in terms of the total mass $M$ and symmetric mass ratio $\eta$ to determine the characteristics of the merging components. These rates can be obtained by integrating over the corresponding variables:
\begin{align}
    \frac{d\mathcal{R}}{dM} & = \int_0^{1/4} \frac{d\mathcal{R}}{dm_1dm_2} \left| \frac{\partial(m_1, m_2)}{\partial(M, \eta)}\right| d\eta\,,
    \\[1ex]
    \frac{d\mathcal{R}}{d\eta} & = \int_0^\infty \frac{d\mathcal{R}}{dm_1dm_2} \left| \frac{\partial(m_1, m_2)}{\partial(M, \eta)}\right| dM\,,
\end{align}
where the Jacobian is the same as~\eqref{eq:Jac}.

\begin{figure}[htbp]
	\centering
	\includegraphics[width=.45\textwidth]{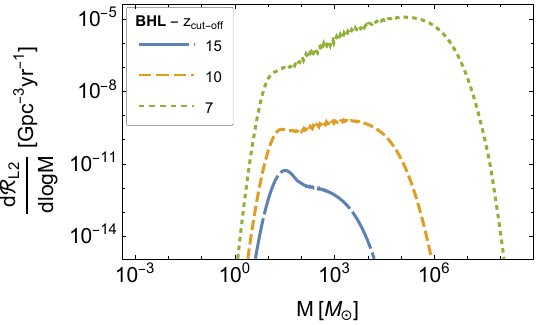}
	\qquad
	\includegraphics[width=.45\textwidth]{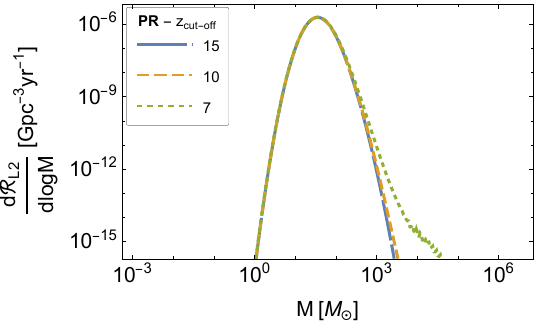}
	
	\includegraphics[width=.45\textwidth]{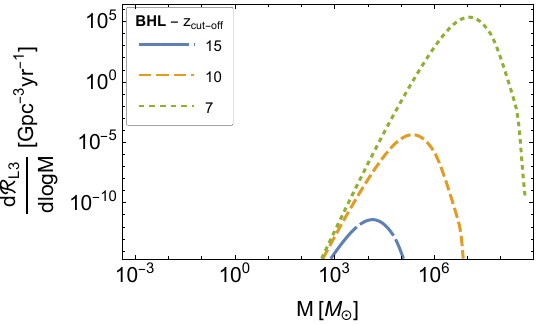}
	\qquad
	\includegraphics[width=.45\textwidth]{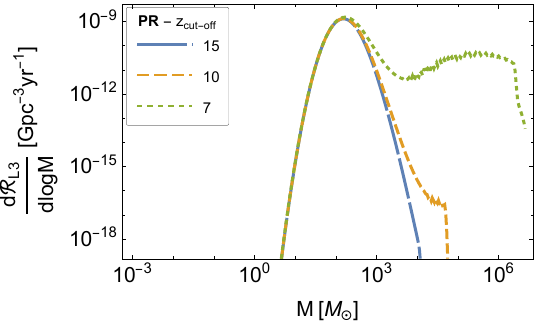}
	\caption{The differential merger rate $\frac{d\mathcal{R}}{d \log_{10} M}$ of binary PBHs as a function of the total mass $M$ at fixed $\langle M\rangle_i = 20\, M_\odot$ and $\gamma = 1$. Rows correspond to the two-body (top) and three-body (bottom) channels, columns represent the BHL (left) and PR (right) accretion models with fixed values $f_i = 10^{-6}$ and $10^{-3}$, respectively.\label{sec:mergerrate:difrate:figM}}
\end{figure}

As hypothesized above (see Section~\ref{sec:population:massfunc}), the reshaping of the binary mass function shifts the most frequent mergers (see Figure~\ref{sec:mergerrate:difrate:figM}) toward heavier masses; furthermore, the presence of an integrand kernels (\ref{R2}, \ref{R3}) with a positive power of the total mass $M$ significantly amplifies this effect. Concurrently, for the PR model, the heavy-mass tail is only slightly raised, whereas for the BHL model, the peak of the distribution itself shifts.

\begin{figure}[htbp]
	\centering
	\includegraphics[width=.45\textwidth]{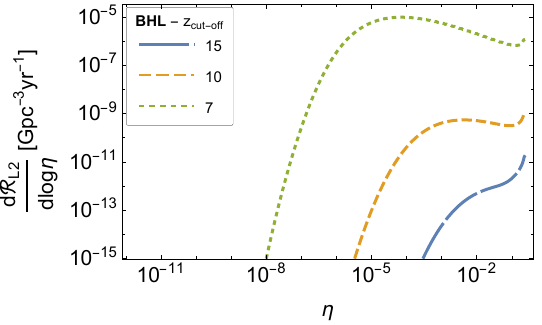}
	\qquad
	\includegraphics[width=.45\textwidth]{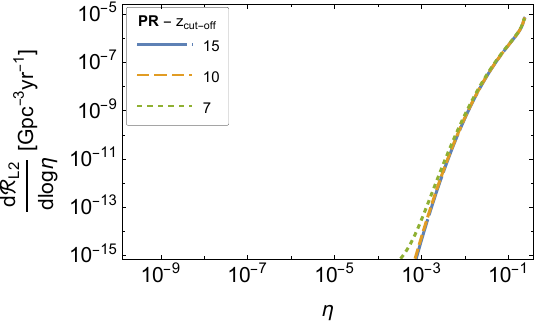}
	
	\includegraphics[width=.45\textwidth]{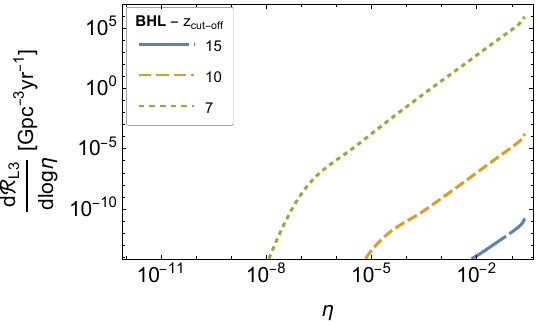}
	\qquad
	\includegraphics[width=.45\textwidth]{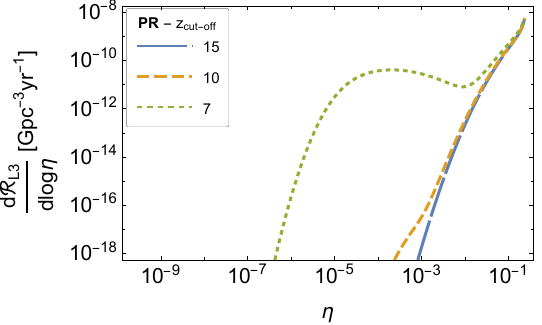}
	\caption{The differential merger rate $\frac{d\mathcal{R}}{d \log_{10} \eta}$ of binary PBHs as a function of the symmetric mass ratio $\eta$ at fixed $\langle M\rangle_i = 20\, M_\odot$ and $\gamma = 1$. Rows correspond to the two-body (top) and three-body (bottom) channels, while columns represent the BHL (left) and PR (right) accretion models with fixed values $f_i = 10^{-6}$ and $10^{-3}$, respectively.\label{sec:mergerrate:difrate:figE}}
\end{figure}

Similarly, mergers of systems with asymmetric component masses become more frequent (see Figure~\ref{sec:mergerrate:difrate:figE}), which is further enhanced by the presence of a negative power of the symmetric mass ratio $\eta$ in the integrand kernels (\ref{R2}, \ref{R3}).

\subsection{Constraints on the PBH fraction in DM}
\label{sec:mergerrate:constraint}

The statistical analysis~\cite{2025arXiv250818083T} of BH-BH, NS-NS, and NS-BH mergers, based on observational data from the LVK gravitational-wave detector network, obtain approximations using various models for the evolution of the binary BH merger rate at $z = 0-1.5$. In that study, the B-Spline approximation model is utilized because it is independent of initial assumptions (see Figure 10 in the cited paper). Since the redshift range considered by those authors does not include $z_\text{cut-off}$, the binary PBH merger rates $\mathcal{R}$ in this work are calculated at $z = 1.5$, cutting off accretion at the considered values of $z_\text{cut-off}$.

An upper limit on the possible binary PBH merger rate formed through the late two- nad three-body channel is set at the LVK level. Based on this constraint, a bound on the initial PBH mass fraction in DM $f_i$ can be derived for the investigated range of initial parameters. For this purpose, the equation 
\begin{equation}
    \mathcal{R}(f_i, z = 1.5) = \mathcal{R}_\text{LVK} \approx 10^2\,\text{Gpc}^{-3}\,\text{yr}^{-1}
\end{equation} is solved, where the merger rate corresponds to that of the three-body channel. This choice is justified because the rate of the two-body channel, as demonstrated above, is significantly lower and would consequently yield a less stringent constraint.

%%%%%%%%%%%%%%%%

\begin{figure}[htbp]
	\centering
	\includegraphics[width=.45\textwidth]{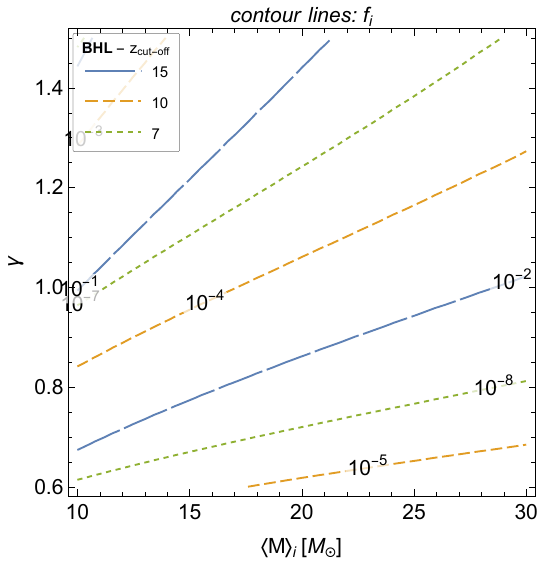}
	\qquad
	\includegraphics[width=.45\textwidth]{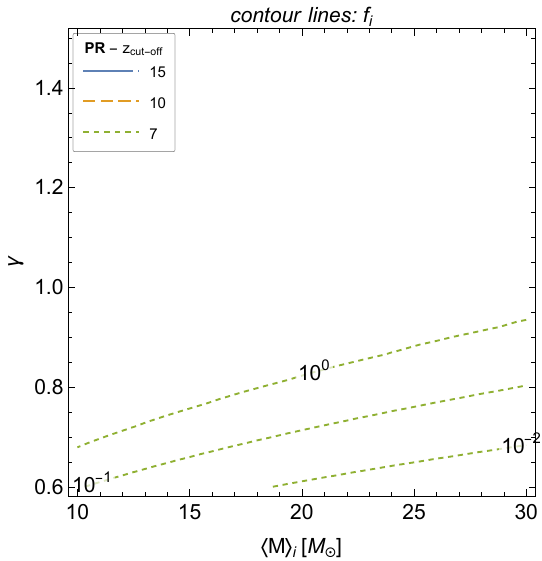}
	\caption{Constraints on the initial PBHs fraction in DM $f_i$ from the LVK merger rate limits as a function of the initial distribution parameters $\langle M\rangle_i, \, \gamma$. The left and right panels correspond to the BHL and PR accretion models, respectively.}
    \label{sec:mergerrate:constraint:fig}
\end{figure}

%In accordance with the obtained solution presented in 
As seen from Figure~\ref{sec:mergerrate:constraint:fig}, 
%, it can be concluded that 
the BHL accretion model imposes strict constraints for any realistic virialization onset redshift. In contrast, the PR model leads to significant constraints only in late-cutoff virialization scenarios.
\section{Merger rate predictions for TianQin space laser interferometer}
\label{sec:tianqin}

This Section presents estimates of the PBHs merger rates in the late-time channel, accounting for preceding post-recombination accretion, for the future space-based interferometer TianQin~\cite{2026LRR....29....1L}, whose sensitivity curve is similar to that of LISA.
%band is optimized for potential intermediate-mass black holes (IMBHs). 

As in the previous Section~\ref{sec:mergerrate}, the initial PBH mass fractions in DM $f_i$ are fixed at $10^{-6}$ and $10^{-3}$ for the BHL and PR accretion models, respectively.

\subsection{TianQin detectability window}
\label{sec:tianqin:detect}

The detectability of merging BHs with masses $m_1$ and $m_2$ at redshift $z$ is determined by the orbital parameters and is limited by two primary factors: the position noise and the residual acceleration noise of the satellites. Accounting for these constraints and averaging over the sky locations and source orientations, the signal-to-noise ratio (SNR) can be defined as \cite{2019PhRvD..99l3002F}:
\begin{equation}
	SNR = \frac{(\mathcal{M} (1+z))^{5/6}}{\sqrt{10}\pi^{2/3} D_L}\sqrt{\int_{\nu_\text{in}}^{\nu_\text{fin}} \frac{\nu^{-7/3}}{S_n(\nu)} \, d\nu}
\end{equation}
where the integration limits are specified as $\nu_\text{in} = \max(\nu_\text{low}, \nu_\text{obs})$ and $\nu_\text{fin} = \min(\nu_\text{end}, \nu_\text{ISCO})$. The corresponding numerical values and analytical formulas are summarized in Table~\ref{sec:tianqin:detect:tab} below (the observation period $T_\text{obs}$ is chosen to be three months).
\begin{table}[htbp]
    \centering
    \begin{tabular}{l|l}
    \hline
    Lower cutoff & \added{$\nu_{\text{low}} = 10^{-4}\,\text{Hz}$} \\ \hline
    Observation time & $\nu_{\text{obs}} = 4.15 \times 10^{-5}\,\text{Hz} \times \left(\frac{\mathcal{M}(1+z)}{10^6 M_\odot}\right)^{-5/8} \left(\frac{T_{\text{obs}}}{1\,\text{yr}}\right)^{-3/8}$ \\ \hline
    Upper cutoff & $\nu_{\text{end}} = 1\,\text{Hz}$ \\ \hline
    Innermost stable circular orbit (ISCO) & $\nu_{\text{ISCO}} = \frac{c^3}{6^{3/2}\pi G M(1+z)}$ \\ \hline
    \end{tabular}
    \caption{Frequency constraints for the TianQin detector \added{\cite{2025CQGra..42q3001L}}.}
    \label{sec:tianqin:detect:tab}
\end{table}

Additionally, the one-sided power spectral density (PSD) of the TianQin noise, $S_n(\nu)$, is determined by the instrumental specifications: the arm length $L_0 = 1.73 \times 10^5\,\text{km}$, the position noise $S_x = 10^{-24}\,\text{m}^2\,\text{Hz}^{-1}$ and the residual acceleration noise $S_a = 10^{-30}\,\text{m}^2\,\text{s}^{-4}\,\text{Hz}^{-1}$:
\begin{equation}
	S_n(\nu) = \frac{S_x}{L_0^2}+\frac{S_a}{4\pi^4 L_0^2 \nu^4}\left(1+\frac{10^{-4} \,\text{Hz}}{\nu}\right)\,.
\end{equation}

The chirp mass (the mass directly measured from GW observations) is expressed in terms of the component masses $m_1$ and $m_2$ 
\begin{equation}
	\mathcal{M} = \frac{(m_1 m_2)^{3/5}}{(m_1+m_2)^{1/5}}\,.
\end{equation}

The luminosity distance $D_L(z)$ as a function of the cosmological redshift $z$ (with the cosmological parameters $\Omega_M$, $\Omega_\Lambda$, and $H_0$ provided above in Section~\ref{sec:accretion:BHL}) is given by:
\begin{equation}
	D_L (z) = \frac{c\,(1+z)}{H_0}\int_0^z \frac{dz'}{\sqrt{\Omega_M (1+z')^3+\Omega_\Lambda}}
\end{equation}

The signal-to-noise ratio threshold is set at $\text{SNR}_\text{det} = 10$, corresponding to a confident event detection. Consequently, this yields an implicit equation defining the surface $\text{SNR}_\text{det} = \text{SNR}(z_\text{lim},\, m_1, \, m_2)$, which can be solved numerically to obtain the dependence $z_\text{lim} = z_\text{lim}(m_1,\,m_2)$.

\subsection{Annual merger rates}
\label{sec:tianqin:rate}

To estimate the event rate per year, the merger rate must be integrated over the comoving volume assuming isotropic distribution of sources across all sky directions, with the radial integration limit bounded by $z_\text{lim}$ to ensure a signal-to-noise ratio above $\text{SNR}_\text{obs}$:

\begin{equation}
\label{eq:dNdt}
	\dot{N} = \int_0^{z_\text{lim}(m_1,\,m_2)} dD_l(z) \iint dm_1\, dm_2\, \frac{d \mathcal{R}_L(m_1, m_2, z)}{dm_1 dm_2}\frac{4 \pi D_L^2(z)}{1+z}
\end{equation}
where the factor of $(1+z)^{-1}$ accounts for the cosmological time dilation between the source rest frame and the detector frame.

\begin{figure}[htbp]
	\centering
	\includegraphics[width=.45\textwidth]{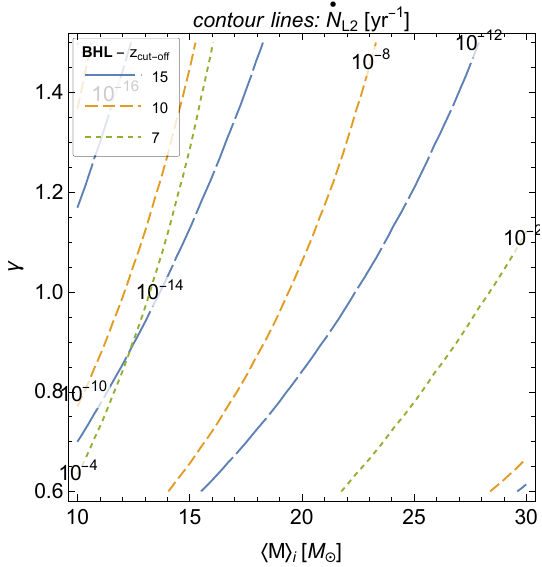}
	\qquad
	\includegraphics[width=.45\textwidth]{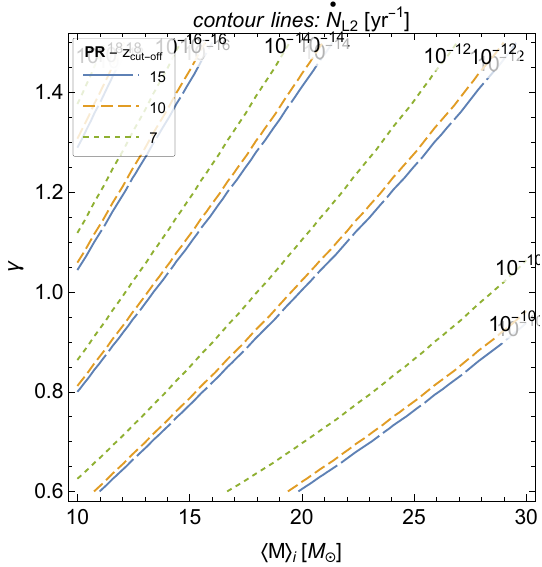}
	
	\includegraphics[width=.45\textwidth]{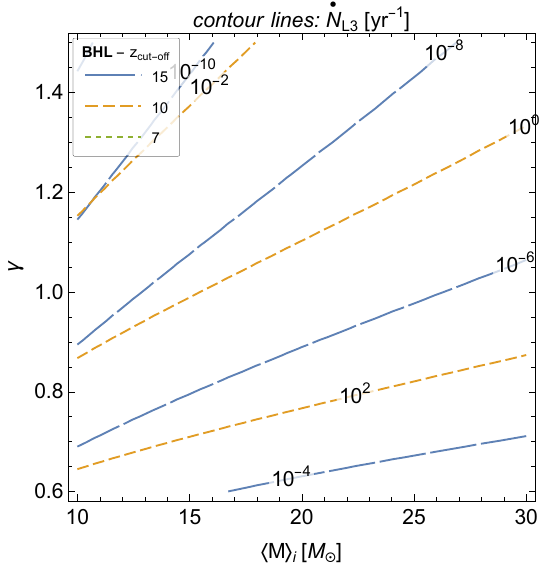}
	\qquad
	\includegraphics[width=.45\textwidth]{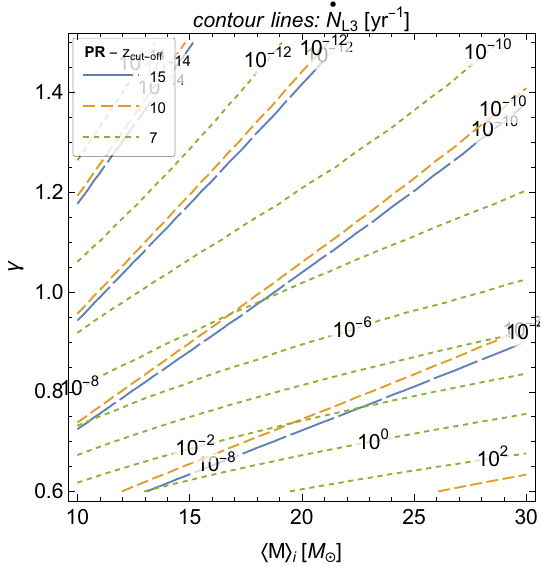}
	
	\caption{The predicted annual merger rate $\dot{N}$ of binary PBHs as a function of the initial distribution parameters $\langle M\rangle_i,$ and $\gamma$. Rows correspond to the two-body (top) and three-body (bottom) channels, while columns represent the BHL (left) and PR (right) accretion models with fixed values $f_i = 10^{-6}$ and $10^{-3}$, respectively.\label{sec:tianqin:rate:fig}}
\end{figure}

According to the integration results (see Figure~\ref{sec:tianqin:rate:fig}), the two-body channel remains undetectable by the TianQin detector for both accretion models. In contrast, the three-body channel becomes detectable within the BHL model; moreover, it yields an unphysically high event rate in the case of a late $(z_\text{cut-off}<10)$ post-recombination accretion cutoff, whereas it remains undetectable for the PR model.

\subsection{Profiles of the annual merger rates}
\label{sec:tianqin:difrate}

The predicted differential annual event rates are also calculated in terms of the total mass $M$ and symmetric mass $\eta$ ratio variables, accounting for post-recombination accretion, for the TianQin detector. These rates are obtained by integration over a single variable ($\eta$ and $M$, respectively) utilizing the Jacobian from~\eqref{eq:Jac} in a similar manner to~\eqref{eq:dNdt}.
\begin{figure}[htbp]
\includegraphics[width=.45\textwidth]{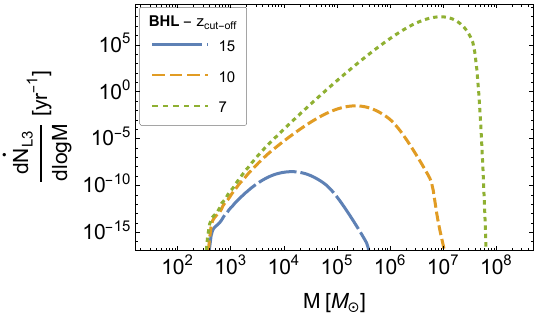}
	\qquad
	\includegraphics[width=.45\textwidth]{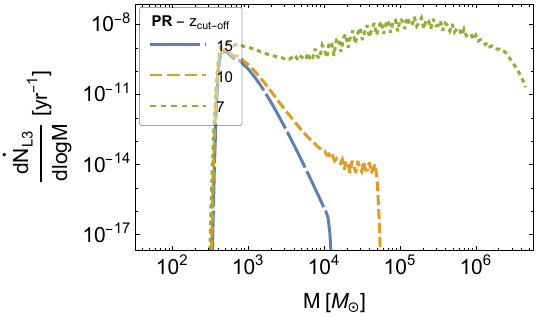}
    
    \includegraphics[width=.45\textwidth]{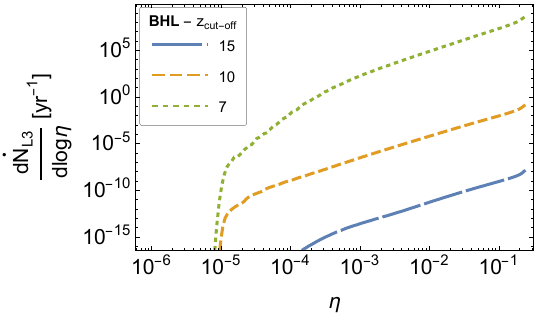}
	\qquad
	\includegraphics[width=.45\textwidth]{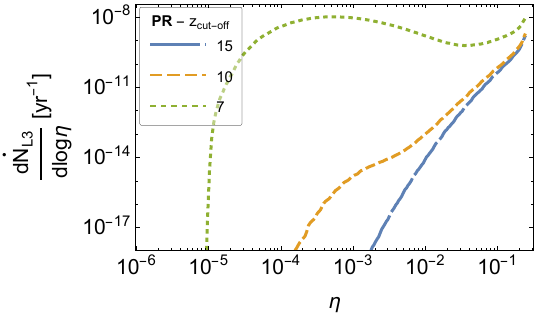}
	\caption{The predicted annual profiles for late three-body channel at fixed $\langle M\rangle_i = 20\, M_\odot$ and $\gamma = 1$. Rows correspond to the functions of total mass $M$ (top) and symmetric mass ratio $\eta$ (bottom), while columns represent the BHL (left) and PR (right) accretion models with fixed values $f_i = 10^{-6}$ and $10^{-3}$, respectively.
    \label{sec:tianqin:difrate:fig}}
\end{figure}

Since the late two-body channel remains undetectable (see Figure~\ref{sec:tianqin:rate:fig}) by the TianQin interferometer, annual merger profiles are presented for the three-body channel only. The obtained plots (see Figure~\ref{sec:tianqin:difrate:fig}) are qualitatively similar in shape to those of the differential rates (see Figures~\ref{sec:mergerrate:difrate:figM} and~\ref{sec:mergerrate:difrate:figE}). The integration over the detector sensitivity curve acts as a filter, cutting off the astrophysical mass and extremely low mass ratios, while emphasizing the heavy-mass tail due to the peak detector sensitivity in this domain.

Thus, the calculated differential rates suggest that if the late-time merger channel is detected by the TianQin interferometer (for late virialization $z_\text{cut-off}$ and a relatively high initial PBH mass fraction in DM $f_i$), the slope of the symmetric mass ratio distribution near the equal-mass limit will be less steep (indicating more frequent asymmetric merger events) than in the case where post-recombination accretion is insignificant.

\section{Discussion}
\label{sec:discussion}

In this section, the primary sources of uncertainty inherent in the utilized models and approximations are outlined, and their impact on the final results of our study is estimated. A cautious extrapolation of the results to alternative initial parameters and detectors, which are not discussed in detail here, is also performed.

\subsection{Treatment of accretion}

The treatment of accretion process strongly modifies the rate of mergers (see the structure of integrand kernels \ref{R2} and \ref{R3}):
\begin{itemize}
    \item \textit{DM halo virialization timescale} \\
    In the present study, we assume instantaneous VDMHs formation, resulting in a simultaneous termination of accretion across the entire PBH population. In reality, a finite timescale is required for VDMH formation, as demonstrated, e.g., in~\cite{2008ApJ...680..829R}. For more precise evaluations of PBH binary formation rates, joint treatment of both accretion and merger processes at low redshifts is required. In the first approximation, these processes can be calculated independently by introducing the fraction of virialized PBH relative to the total (initially isolated) PBH population.
    \item \textit{Velocity profile} \\
    In the numerical calculations of the BHL and PR accretion models we have employed the ROM08 velocity profile. This profile was derived by Ricotti~\cite{2008ApJ...680..829R} from a detailed CMB analysis. Alternatively, a simplified SPIK20 velocity profile is used in other studies~\cite{2020PhRvR...2b3204S}, where the PBH and sound velocities are parameterized as:
    \begin{equation}
    \label{e:veffz}
    	\begin{aligned}
    		v_\text{PBH} &= 30 \min\left[\frac{1+z}{1000},\, 1\right] \quad \text{km}\,\text{s}^{-1}, \\
    		c_s          &= 5.7 \sqrt{\frac{1+z}{1000}} \quad \text{km}\,\text{s}^{-1}.
    	\end{aligned}
    \end{equation}
    As noted in \cite{2025JCAP...08..006J}, the use of the SPIK20 profile instead of ROM08 enhances the accretion rate in the BHL model and suppresses it in the PR model. Consequently, the final results for the accretion-driven mass growth can differ by up to several orders of magnitude.
    \item \textit{Sound Speed Contrast} \\
    The range of possible values for the sound speed contrast, $\frac{c_s^\text{in}}{c_s} = 10\text{--}50$, within the PR accretion model has been addressed in Ref.~\cite{2025JCAP...08..006J}. The choice of a smaller value results in an order-of-magnitude enhancement of the accretion rate; conversely, a larger value leads to its order-of-magnitude suppression.
    \item \textit{Angular momentum distribution index} \\
    The distribution of orbital angular momenta for PBH binaries employed in the Raidal model is taken as a power-law function:
    \begin{equation}
    	\frac{dP}{dj}  = \gamma_a j^{\gamma_a - 1},
    \end{equation}
    where $\gamma_a \in \left[1,\,2\right]$. As noted in~\cite{2024arXiv240408416R}, the complete thermalization ($\gamma_a = 2$) is not achieved in self-gravitating many-body systems. The use of a lower index $\gamma_a$ results in an enhancement of PBH merger rates by up to two orders of magnitude.
\end{itemize}

In addition, a detailed treatment of the transition from the Bondi accretion regime to the Eddington regime, and subsequently to the super-Eddington regime, must be conducted. The necessity of refining the PR accretion model by introducing the $\lambda$ factor analogous to the BHL framework is also suggested. This will allow the influence of external cosmological factors to be accounted for and is expected to suppress the accretion rate, primarily for heavy PBHs. Such a modification should ensure the correctness of the regime transitions.

\subsection{Notes on disk accretion}

In the classical BHL accretion setup, no accretion disk is expected to form around a BH moving through a homogeneous medium \cite{2025ApJ...979...61T}, unless density or velocity gradients are present on the gravitational radius scale. It is unclear whether such gradients (e.g., caused by IGM turbulence) can be present before the structure formation and reionization. Simple estimates of the requirements for an accretion disk to form around an accreting BH from IGM can be done in a way similar to considerations in ISM \cite{2005A&A...440..223B}.  For a disk to form around a BH, the specific angular momentum of the captured particle should exceed that at the last stable circular orbit around a non-rotating BH, $l>\sqrt{3}r_g c$, where $r_g=\frac{2GM}{c^2}$ is the BH gravitational radius. The angular momentum gravitationally captured by a BH moving with velocity $v$ through inhomogeneous medium with sound speed $c_s$ as derived from numerical simulations $l\sim 0.1 l_m$ \cite{1999A&A...346..861R}\footnote{More recent 3D-simulations of BHL accretion \cite{2019MNRAS.488.5162X} suggest even less efficient angular momentum capture suggesting a more prone accretion to instabilities in the case of small transverse density gradients.} , where $l_m=\beta\frac{1}{4}v_\text{eff}(2r_B)$ is the specific angular momentum of captured matter available from the accreting cylinder \cite{1975A&A....39..185I} and $\beta=\left(\frac{\Delta v}{v}, \frac{\Delta \rho}{\rho}\right)$ is the velocity and/or density fluctuations on scale $2r_B$. For the assumed Poissonian distribution of PBH $f_\text{PBH i}\ll 1$ no velocity fluctuations are expected, and short-wavelength baryonic density perturbations can be taken from CMB measurements as $\beta=\frac{\Delta\rho}{\rho}\Delta_\mathcal{R}\sim 5\times 10^{-5}\left(\frac{1100}{1+z}\right)$ with flat spectrum. Thus the condition for spherical BHL accretion reads
\begin{equation}
    \label{e:spherical}
    \beta\approx  5\times 10^{-5}\left(\frac{1100}{1+z}\right)<40\sqrt{6}\frac{v_\text{eff}}{c}
\end{equation}
After recombination, PBH peculiar velocity decreases as $v_\text{eff}\sim 30\,\text{km/s}\frac{1000}{1+z}$  (see above Equation (\ref{e:veffz})), thus the condition for spherical accretion becomes 
\begin{equation}
    (1+z)_{sph}\gtrsim 80\,.
\end{equation}
(Note that less efficient angular momentum capture than 10\% we assumed will decrease this estimate accordingly.) 
This estimate is very close to our use of ROM08 prescriptions for luminosity-weighted effective velocities of PBH fro disk accretion  (see Figure 2, curve $\langle v_\text{eff}\rangle_B$ in that paper). Clearly, the possible PBH clustering \cite{2017PhRvD..96h3524P} renders disk accretion onto PBH more likely immediately after recombination, and the use of PR prescription for accretion rate instead of BHL can be more appropriate. However, here we have ignored PBH clustering because of
tiny PBH fraction in CDM $f_i\ll 1$ we consider.

\subsection{Extrapolations of the results}

Furthermore, our results can be cautiously extrapolated as follows:
\begin{itemize}
    \item \textit{Power-law PBH mass function}\\
    In modern literature, a power-law PBH mass function is also frequently considered. Although computations for this specific distribution are not performed in the present work, the corresponding conclusions are expected to be qualitatively similar to our findings, yet more pronounced due to slower decay of the heavy tails in the power-law mass distribution compared to log-normal profiles. Consequently, an even higher fraction of heavy PBHs is expected to be observed in the population post-accretion. This mechanism should result in a stronger enhancement of the merger rates and a more significant flattening of their distributions within the late-time channel.
    \item \textit{Evolution of PBH mass distribution at radiation-dominated stage}\\
    The PBH mass function can change already at the RD stage. Recent numerical GR calculations of accretion onto PBH at the radiation-dominated stage \cite{2025PhRvD.112l3540D} suggest that there can be saturation of the PBH mass growth. This effect can affect the PBH mass increase at the radiation-dominated stage carried out in the Newtonian approximation \cite{2017JHEAp..13...22R} and should be investigated further.

    \item \textit{Other space-based GW interferometers}\\
    In this work, the sensitivity curve of the TianQin detector is exclusively considered; however, while the characteristics of other space-based interferometers differ in detail, their sensitive black hole mass ranges remain similar. Since the computed differential merger (section~\ref{sec:mergerrate:difrate}) rates are intrinsic and completely independent of the instrument, whereas the detector itself acts as an instrumental mask, qualitatively similar features in the differential binary PBH merger rates are expected for other interferometers: a shift in the peak rate toward higher total masses and a notable flattening with respect to the symmetric mass ratio.
    
    \item \textit{Upper bound of the initial PBH mass heavy tail}\\
    In our calculations the right tail of the initial PBH mass distributions is truncated at $10^{4}M_{\odot}$. A qualitative analysis of modifications arising from the potential existence of PBHs with initial masses exceeding this limit is of interest. In this case, an extension of the integration mass interval would lead to a minor increase in the PBH fraction $f_z$ at the redshift $z_{\text{cut-off}}$. As the estimated differential merger rates  peak before the upper bound of the accreted masses is reached (see Figure~\ref{sec:mergerrate:difrate:figM}), the existence of PBHs with larger initial masses would extend the differential rate curve below the peak level \added{(which holds for the considered parameters $\langle M\rangle_i = 20 M_\odot$ and $\gamma = 1$, and is expected to remain qualitatively valid for other cases)}. Consequently, only a minor enhancement in the annual merger rates is expected due to expansion of the integration mass interval and increase in the accretion-induced PBH fraction $f_z$.
\end{itemize}

\section{Conclusion}
\label{sec:conclusion}

In this study, we obtained the following results that characterize the merger statistics of PBHs in virialized VDMHs, accounting for the evolution of their population parameters driven by post-recombination accretion in the Bondi--Hoyle--Littleton and Park--Ricotti  models:
\begin{itemize}
    \item Post-recombination accretion yields larger enhancement factors $\mathcal{E}$ for the three-body channel of binary PBH formation than for the two-body one. The BHL accretion model has a  significantly greater impact on the binary PBH merger rate compared to the PR model.
    \item Estimates of the merger rate enhancement factor $\mathcal{E}$ have been obtained, comparing scenarios with preceding accretion against those without it. This factor is mostly affected by 
    %increases under the following variations of the initial mass function parameters:
    an increase in the initial mean PBH mass $\langle M \rangle_i$, a decrease in the mass distribution shapness $\gamma$, and shift the accretion cutoff $z_\text{cut-off}$ to lower redshifts. Remarkably, the latter parameter has a significantly greater influence on the enhancement than the initial parameters of the PBH mass function.
    \item Constraints on the initial fraction of PBHs (initially unassociated) in DM $f_i$ at the recombination epoch have been derived. The PR accretion model, which precedes the mergers, does not impose new constraints across the broad range of initial mass function parameters considered in this work. Conversely, the BHL accretion model leads to stringent constraints; thus, this accretion model is physically viable only under an extremely low initial fraction of isolated PBHs $f_i$ or with a sufficiently early onset of VDMHs formation ($z_\text{cut-off}$).
    \item The predicted annual event rate for the future space-based gravitational-wave  interferometer TianQin has been evaluated. Within the BHL model, the feasibility of registering events in the late-time three-body channel has been demonstrated; previously, without accounting for the mass function reshaping driven by accretion, this channel was considered suppressed and inaccessible to detections.
    \item The reshaping of the differential merger rate profiles driven by preceding accretion has been demonstrated, featuring a peak shift (or tail raising) toward heavier masses, as well as a less steep slope near the equal-mass limit, which implies an increased fraction of highly asymmetric merger events. These effects will manifest themselves by the TianQin interferometer in the event of a successful detection of the binary BH mergers formed in the late-time channel.
\end{itemize}

Overall, our results highlight the necessity of accounting for post-recombination accretion (especially in relatively efficient accretion scenarios) in the analysis of the PBH merger statistics within the late-time channel.

\acknowledgments
The authors thank \added{the anonymous referee for useful comments} and Prof. M. Pshirkov for discussions.
\added{The study was conducted under the state assignment of Lomonosov Moscow State University.}\\

This is the Accepted Manuscript version of an article accepted for publication in \textit{Journal of Cosmology and Astroparticle Physics}. IOP Publishing Ltd is not responsible for any errors or omissions in this version of the manuscript or any version derived from it. The Version of Record is available online at \href{https://doi.org/10.1088/1475-7516/2026/11/005}{https://doi.org/10.1088/1475-7516/2026/11/005}.

\bibliographystyle{JHEP}
\bibliography{biblio.bib}

\end{document}